\documentclass[aps,pre,reprint,superscriptaddress,amsmath,amssymb]{revtex4-2}

\usepackage{graphicx}
\usepackage{xcolor}
\usepackage{dcolumn}
\usepackage{bm}
\usepackage{caption}
\newcommand{\nc}{\newcommand}           
\nc{\beq}     {\begin{eqnarray}}
\nc{\eeq}    {\end{eqnarray}}
\nc{\nn}  {\\\nonumber}
\nc{\D} {\Delta}
\nc{\red}[1]    {\textcolor{red}{#1}}  
\nc{\blue}[1]   {\textcolor{blue}{#1}}  
\nc{\green}[1]   {\textcolor{green}{#1}}  

\begin{document}


\title{Collective Signatures of Pionic Correlations at High Momentum Transfer}

\author{Hiroshi \textsc{Toki}}
\email{tokihiro12@gmail.com}
\affiliation{Research Center for Nuclear Physics (RCNP), The University of Osaka, Ibaraki, Osaka 567-0047, Japan}

\author{Junki \textsc{Tanaka}}
\email{junki@rcnp.osaka-u.ac.jp}
\thanks{Corresponding author}
\affiliation{Research Center for Nuclear Physics (RCNP), The University of Osaka, Ibaraki, Osaka 567-0047, Japan}
\affiliation{Graduate School of Science, The University of Osaka, Toyonaka, Osaka 560-0043, Japan}

\date{\today}


\begin{abstract}

High-momentum-transfer reactions have provided experimental evidence for enhanced pionic correlations in nuclei, yet the microscopic connection between the low-energy pionic response and the $\D$-hole excitation spectrum remains unresolved. Here we identify hybrid pion-delta collective modes that emerge through the coherent coupling of nucleon-hole and delta-hole excitations. By evaluating the full delta-hole Lindhard function within the random-phase approximation, we find a robust low-energy hybrid pion-delta mode and a higher-energy collective delta-hole mode. The low-energy mode is strongly enhanced as the Landau-Migdal parameter is reduced, exhibiting behavior characteristic of a precursor to pion condensation. We further show that these modes may be probed through high-momentum-transfer charged-pion production reactions. Because the same collective pion-delta function governs both the response and the RPA correlation energy, our results connect experimentally observable collective excitations with microscopic many-body correlations.

\end{abstract}

\maketitle


\section{Introduction}

Collective spin-isospin excitations generated by pion exchange have long been recognized as a central aspect of nuclear many-body dynamics. Beginning with the pioneering work of Migdal, pion condensation and its precursor phenomena were extensively investigated in nuclear matter~\cite{Migdal1978}. Subsequent studies of finite nuclei by Toki and Weise, and independently by Delorme \textit{et al.}, demonstrated that strong pionic correlations produce enhanced pion-like excitations at large momentum transfer, providing signatures of precursor phenomena of pion condensation~\cite{Toki1979, Delorme1980}. These early studies established that the momentum dependence of the pion-exchange interaction can generate collective spin-isospin modes well below the pion-condensation instability.

Although ordinary nuclei are far from the pion-condensed phase, strong pionic correlations persist in finite nuclei and nuclear matter. Owing to the large $\pi N\Delta$ coupling, $\Delta$-hole ($\Delta h$) excitations play an essential role in the spin-isospin response and strongly modify the excitation spectra. In early studies of low-lying pionic spectra, however, the influence of $\Delta h$ excitations was incorporated mainly through simplified approximations, retaining only the real part of the $\Delta h$ self-energy~\cite{Toki1980}. Their explicit role in high-energy excitation spectra was later analyzed by Chanfray and Ericson~\cite{Chanfray1984}, and subsequent analyses of coherent pion production demonstrated the collective character of the nuclear pionic response~\cite{Delorme1991, Dmitriev1993, Hennino1992, Hennino1993, Kagarlis1997, Boyard2000}. These developments raise a fundamental question: how are the low-energy collective pionic mode and the collective $\Delta h$ excitation dynamically coupled in a microscopic description?

The necessity of explicit $\Delta$ degrees of freedom is also firmly established in modern microscopic nuclear theory. A classic example is the Fujita-Miyazawa three-nucleon force, in which intermediate $\Delta$ excitations generate effective three-body interactions~\cite{Fujita1957}. Microscopic calculations incorporating explicit $N\Delta$ and $\Delta\Delta$ channels further showed that $\Delta$ excitations contribute substantially to nuclear binding and structure~\cite{Wiringa1984, Pieper2001, Horii2014}. 
In parallel, self-consistent treatments of pion propagation began to include both nucleon-hole ($Nh$) and $\Delta h$ excitations together with short-range spin-isospin correlations~\cite{Nishida1995}. Polarization-transfer measurements in quasielastic $(p,n)$ reactions separated the spin-longitudinal and spin-transverse responses within the distorted-wave impulse approximation (DWIA) analysis, revealing a characteristic enhancement of the longitudinal response at momentum transfers $q \simeq 2\text{--}3\,m_\pi$~\cite{Wakasa2004, Kawahigashi2001}. Continuum-RPA analyses based on the $\pi+\rho+g'$ interaction subsequently provided a unified description of the Gamow-Teller and quasielastic spin-longitudinal responses~\cite{Wakasa2005, Wakasa2006}, establishing clear experimental evidence for strong pionic correlations in nuclei.  More recently, pionic correlations, short-range correlations, and nuclear-matter response functions have been connected within microscopic descriptions of the nuclear equation of state~\cite{Chanfray2007, Massot2008, Massot2009, Chanfray2023}. These advances motivate a unified reexamination of collective pion dynamics in which $Nh$ and $\Delta h$ excitations are treated on an equal footing.

In the present work, we investigate the dynamical pionic response at high momentum transfer by going beyond previous approximations: we evaluate the full $\Delta h$ Lindhard function, explicitly performing the complete Fermi-sea integration. This allows us to describe the energy and momentum dependence of $\Delta h$ excitations on the same footing as $Nh$ excitations. By summing ring diagrams to all orders within the random-phase approximation (RPA), we examine the mutual coupling of $Nh$ and $\Delta h$ excitations through the spin-longitudinal interaction and its dependence on short-range correlations.

\begin{figure*}[t]
\centering
\includegraphics[width=0.65\textwidth]{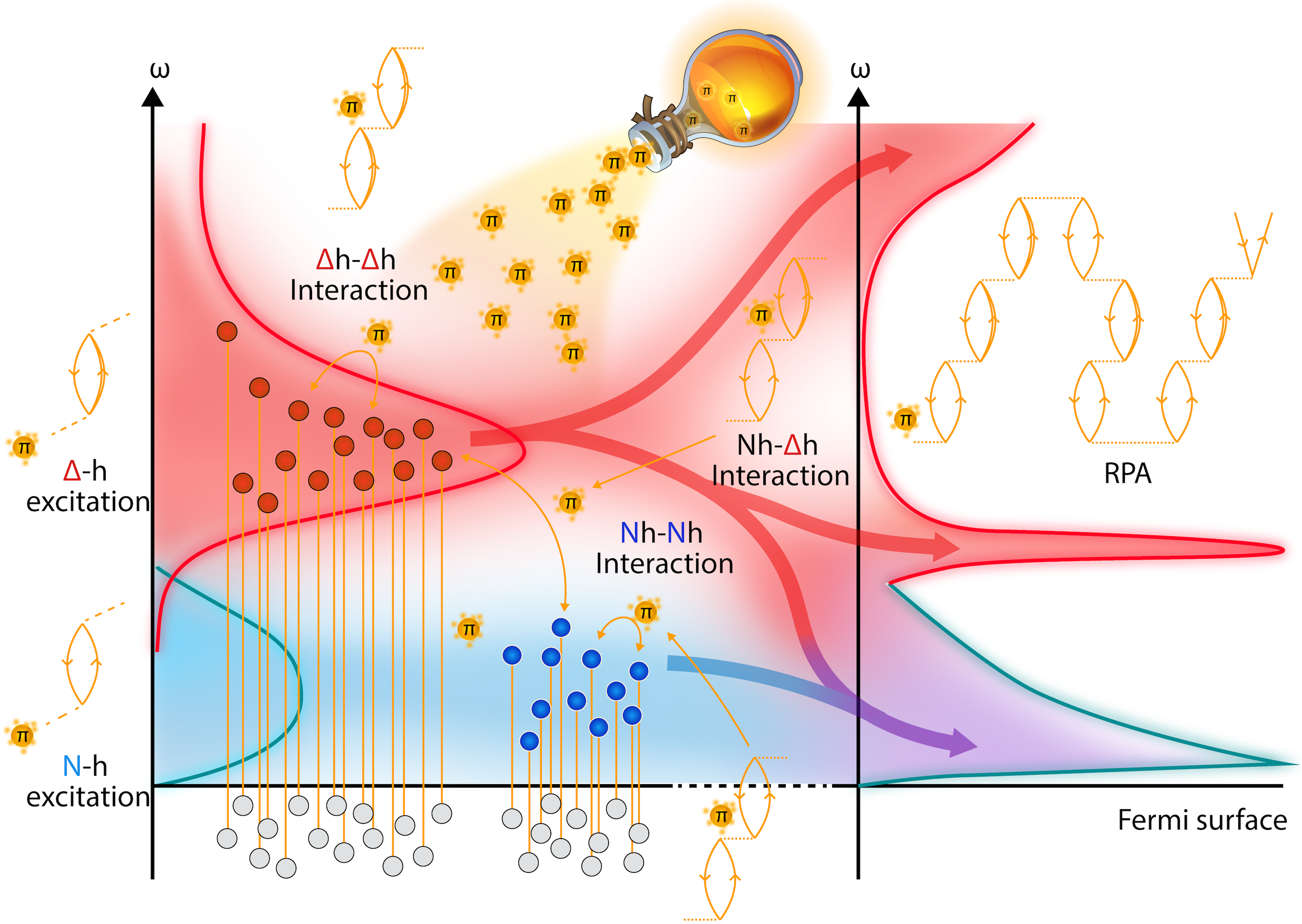}
\caption{Schematic illustration of the collective modes predicted
in the present work.
The free response functions associated with the $Nh$ and
$\Delta h$ excitations are shown on the left
as contributions to the polarization function
$\Pi(\omega,q)$ at $q \sim 3m_{\pi}$ and normal nuclear density.
The pion-exchange interaction couples the $Nh$ and $\Delta h$
excitations through RPA correlations, generating both the
low-energy collective pion mode and the collective $\Delta h$
mode, as represented by the RPA response function
$R(\omega,q)$ on the right.
The vertical axis represents the excitation energy $\omega$,
while the horizontal direction illustrates how pion-mediated
correlations reshape the response.}
\label{fig:schematic}
\end{figure*}

We find that the coupled $Nh$--$\Delta h$ response gives rise to two distinct collective structures: a pronounced low-energy hybrid pion-$\Delta$ mode emerging at momenta of a few pion masses, and a higher-energy collective structure developing in the $\Delta h$ region. As illustrated schematically in Fig.~\ref{fig:schematic}, the full Fermi-sea integration spreads the $\Delta h$ strength over a broad energy range, producing a substantial low-energy tail that strongly mixes with $Nh$ excitations. The low-energy mode is strongly enhanced as the Landau-Migdal parameter $g'$ is reduced, exhibiting behavior characteristic of a precursor to pion condensation.

These predicted collective modes can be experimentally probed through high-momentum-transfer reactions sensitive to the longitudinal spin-isospin response, such as proton-induced charged-pion production $(p, p'\pi^\pm)$ and charge-exchange reactions like $(^3\text{He}, t)$ and $(p,n)$. Crucially, because the same collective pion-$\Delta$ function governs both the dynamical response function and the RPA correlation energy, our results directly link experimentally observable collective excitations to the microscopic mechanism of nuclear binding.

\section{Theoretical Framework}

We formulate a nuclear many-body framework in which pions, nucleons, and $\Delta$ isobars are treated as explicit dynamical degrees of freedom, with nucleon-hole ($Nh$) and $\Delta$-hole ($\Delta h$) excitations incorporated on an equal footing. The resulting framework focuses on the longitudinal spin-isospin response, where pion exchange generates strong correlations through coupling to $N h$ and $\Delta h$ excitations.
To describe these correlations nonperturbatively, we employ the random-phase approximation (RPA) and sum the ring diagrams to all orders. 
This framework provides a unified description of the coupled $N h$ and $\Delta h$ dynamics responsible for the hybrid pion-delta collective modes.

The spin-isospin interaction is described by pion exchange supplemented by a short-range Landau-Migdal interaction~\cite{Migdal1978,Oset1982},
\begin{equation}
V(\omega,\bm{q})=V_\pi(\omega,\bm{q})+V_{\rm LM}(\omega,\bm{q})~.
\end{equation}
In the $NN$ channel, the pion-exchange interaction is given by
\begin{equation}
V_\pi^{NN}(\omega,\boldsymbol{q}) = \left( \frac{f}{m_\pi} \right)^2 \frac{(\boldsymbol{\sigma}_1 \cdot \boldsymbol{q}) (\boldsymbol{\sigma}_2 \cdot \boldsymbol{q})}{\omega^2 - \boldsymbol{q}^2 - m_\pi^2} (\boldsymbol{\tau}_1 \cdot \boldsymbol{\tau}_2),
\end{equation}
where $f$ is the $\pi N N$ coupling constant ($f^2/4\pi \simeq 0.08$). For processes involving $\Delta$ isobars, we replace the nucleon spin/isospin operators $\boldsymbol{\sigma}, \boldsymbol{\tau}$ and the coupling constant $f$ with the transition operators $\boldsymbol{S}, \boldsymbol{T}$ (which connect the $S=1/2$ and $S=3/2$ spaces) and the $\pi N \Delta$ coupling constant $f^*$, adopting $f^*/f=2$~\cite{EricsonWeise1988}:
\begin{equation}
f \to f^*, \quad \boldsymbol{\sigma} \to \boldsymbol{S}, \quad \boldsymbol{\tau} \to \boldsymbol{T}.
\end{equation}
To account for the finite size of hadrons ($N$ and $\Delta$), a monopole form factor is introduced at each pion vertex,
\begin{equation}
F(\omega,\bm{q})= \frac{\Lambda^2-m_\pi^2}{\Lambda^2-\omega^2+\bm{q}^2}~,
\end{equation}
with $\Lambda=1.2$ GeV.

The short-range Landau-Migdal interaction $V_{\mathrm{LM}}$, which represents the residual interaction not carried by the pion, is given by
\begin{equation}
V_{\mathrm{LM}}(\boldsymbol{q}) = \left( \frac{f}{m_\pi} \right)^2 g' (\boldsymbol{\sigma}_1 \cdot \boldsymbol{\sigma}_2) (\boldsymbol{\tau}_1 \cdot \boldsymbol{\tau}_2).
\end{equation}
Crucially, when $\Delta$ degrees of freedom are included, the $g'$ parameter becomes a matrix in the $NN$ and $N\Delta$ space ($g'_{NN}, g'_{N\Delta}, g'_{\Delta\Delta}$).
Following the commonly adopted universal approximation, we assume
\beq
g'_{NN}=g'_{N\Delta}=g'_{\Delta\Delta}=g'~,
\eeq
where values in the range $0.5$-$0.7$ are considered.

The effective interaction for the $N h$ excitations entering the RPA equations is given by
\beq
V(\omega, \bm q)=\left(\frac{f}{m_\pi}\right)^2F(\omega,\bm q)^2{\bm q}^2 D(\omega, \bm q)~.
\eeq
Here,
\begin{equation}
\label{interaction}
D(\omega, \bm q)=
\frac{1}{\omega^2-\bm q^2-m_\pi^2}+\frac{g'}{\bm q^2}~,
\end{equation}
where the first term is the pion propagator, while the second term originates from the Landau-Migdal interaction, which phenomenologically accounts for the short-range repulsion and exchange effects.

The pion self-energy is then written as
\begin{equation}
\Pi(\omega,\bm q)=
-\bm q^2
\left(
\frac{f}{m_\pi}
\right)^2F(\omega,\bm q)^2
\left[
\chi_N(\omega, \bm q)+\chi_\Delta(\omega, \bm q)
\right]~,
\end{equation}
where $\chi_N$ and $\chi_\Delta$ denote the $N h$ and $\Delta h$ susceptibilities, respectively.  With this definition, the different coupling constants ($f$ and $f^*$) as well as the corresponding spin-isospin operators are incorporated into $\chi_N$ and $\chi_\Delta$.

The longitudinal spin-isospin response function is obtained within the RPA as
\begin{equation}
R(\omega,\bm q)=
\frac{\Pi(\omega,\bm q)}
{1-D(\omega,\bm q)\Pi(\omega,\bm q)}~.
\label{response}
\end{equation}
We define
\begin{equation}
{\cal C}_{\pi\Delta}(\omega,\bm q)
=1-D(\omega,\bm q)\Pi(\omega,\bm q)~,
\label{cpid}
\end{equation}
which we refer to as the collective pion-delta function. Collective modes are identified by the condition
\beq
\mathrm{Re}\,{\cal C}_{\pi\Delta}(\omega,\bm q)=0~,
\eeq
whereas the imaginary part determines the damping width of the modes. The same collective pion-delta function governs both the collective excitations and the RPA correlation energy, establishing a direct connection between excited states and the nuclear ground state \cite{Gell-Mann1957, Fetter2003}.

The $N h$ susceptibility is described by the Lindhard function for a free Fermi gas \cite{Fetter2003}. 
\begin{equation}
\chi_N(\omega,q)=
\frac{N(0)}{2}
\left[
1
-\frac{1}{4x}u(y-x)
+\frac{1}{4x}u(y+x)
\right]~,
\end{equation}
with
\begin{equation}
x=\frac{q}{2k_F}~,
\qquad
y=\frac{M_N\omega}{k_F q}~,
\end{equation}
\begin{equation}
u(z)=
(1-z^2)
\ln
\left(
\frac{z+1}{z-1}
\right)~,
\end{equation}
where $z$ is regarded as a complex variable.  The logarithm is defined by the retarded prescription,
\beq
\ln \left(\frac{z+1}{z-1}\right)
=
\ln \left|\frac{z+1}{z-1}\right|-i\pi,
~~~ -1<\mathrm{Re}\,z<1~,
\eeq
which uniquely determines the analytic continuation of the Lindhard function.  The resulting imaginary part is identical to that obtained by explicit evaluation of the momentum integrals \cite{Fetter2003}. Here and throughout this paper, $q\equiv|\bm{q}|$ denotes the magnitude of the three-momentum.
Furthermore,
\begin{equation}
N(0)=\frac{2M_N k_F}{\pi^2}~,
\end{equation}
denotes the density of states at the Fermi surface including both spin and isospin degeneracies. The only difference from the electron-gas Lindhard function is the additional isospin degeneracy factor.

The inclusion of $\Delta$ excitations is essential because of the large $\pi N\Delta$ coupling. Unlike many earlier treatments, in which the $\Delta h$ excitation was represented by a single excitation energy, the present formulation evaluates the full $\Delta h$ Lindhard function by explicitly carrying out the Fermi-sea integration.  To the best of our knowledge, a complete derivation of the $\D h$ Lindhard function has not been explicitly documented in the literature. For completeness, we present the derivation in this Supplemental Material.

The $\Delta h$ susceptibility is given by
\beq
   && \chi_\Delta(\omega,\bm q) = -\left(\frac{2}{3}\right)^2 \left( \frac{f^*}{f} \right)^2 \times 4\int \frac{d^3k}{(2\pi)^3}  \nn&&\left[ \frac{n(\bm{k})}{\omega - (\varepsilon_\D(\bm{k}+\bm{q}) - \varepsilon(\bm{k})) + i\Gamma_\D(\omega)/2} + (\omega \to -\omega) \right] ~.
\eeq
Here, the factor $(2/3)^2$ originates from the spin- and isospin-transition matrix elements, while the factor 4 accounts for the spin and isospin degeneracies of the nucleon.  
For convenience, the difference between the
$\pi N\Delta$
and
$\pi NN$
coupling constants is absorbed into the definition of
$\chi_\Delta(\omega,\bm q)$,
so that the $N h$ and
$\Delta h$ susceptibilities can be treated on the same footing.

After carrying out the Fermi-sea integration analytically, the $\Delta h$ susceptibility can be written in the compact form
(see the detailed derivation in this Supplemental Material)
\begin{equation}
\label{chid}
\chi_\Delta(\omega,q)=
-\frac{4}{9}
\left(
\frac{f^*}{f}
\right)^2 \frac{M_\D}{M_N}
\frac{N(0)}{2}
\left[
\Phi(\omega,q)
+
\Phi(-\omega,q)
\right],
\end{equation}
where
\begin{equation}
\Phi(\omega,q)=
\frac{1}{4x}
\left[
(1-\bar y^2)
\ln
\left|
\frac{\bar y+1}
{\bar y-1}
\right|
+
2\bar y
\right]
\end{equation}
with
\begin{equation}
\bar y=
\frac{
M_\Delta(\tilde\omega-\tilde\omega_\Delta)-q^2/2
}
{k_F q}~.
\end{equation}

A finite $\Delta$ width is incorporated through
\beq
\tilde\omega
=\omega+i\frac{\Gamma_\D(\omega)}{2}~,
\eeq
and
\beq
\tilde\omega_\Delta=
\omega_\Delta
+
\frac{k_F^2}{2}
\left(
\frac{1}{M_\Delta}-
\frac{1}{M_N}
\right)~,
\eeq
includes the recoil correction arising from the mass difference between the nucleon and the $\Delta$ resonance.

The finite lifetime of the $\Delta$ resonance is taken into account
through an energy-dependent width,
\beq
\Gamma_\Delta(\omega)=
\Gamma_0
\left(
\frac{\omega}{\omega_\Delta}
\right)^2~ ,
\eeq
where $\Gamma_0=50$ MeV is adopted in the present calculation.
Details of the parametrization are given in the Supplemental Material.
The dominant effect originates from the full
$\Delta h$ Lindhard function rather than from the particular choice
of the $\Delta$ width.

\section{Results and Discussion}
\begin{figure}[h]
\centering
\includegraphics[width=\linewidth]{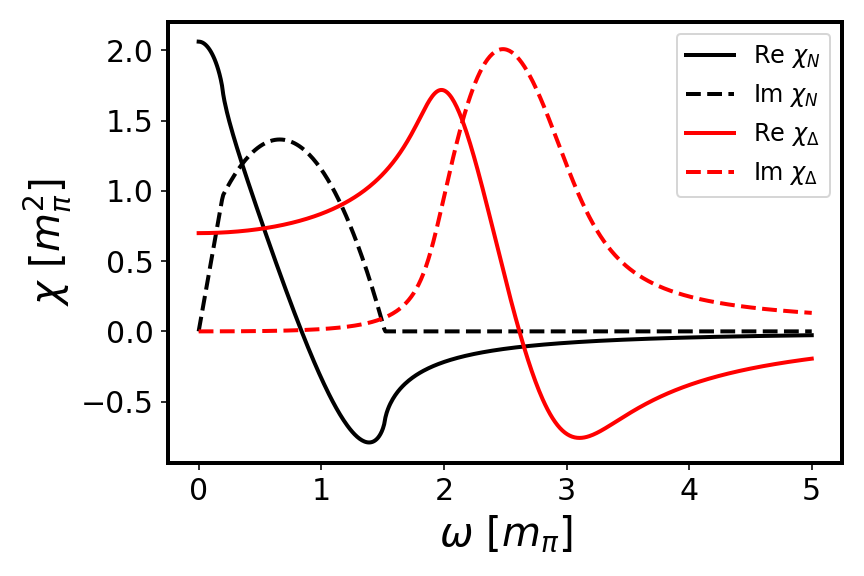}
\vspace{-3mm}
\caption{
Real and imaginary parts of the $Nh$ and $\Delta h$
susceptibilities, $\chi_N(\omega,q)$ (black) and
$\chi_\Delta(\omega,q)$ (red), respectively,
at normal nuclear matter density for $q=3m_\pi$.
The $Nh$ response is concentrated in the low-energy
region, while the imaginary part of the $\Delta h$
susceptibility exhibits slightly greater strength
owing to the strong $\pi N\Delta$ coupling.
The full $\Delta h$ Lindhard function, including
the Fermi-sea integration and the finite $\Delta$
width, distributes the $\Delta h$ strength over
a broad energy range, producing a substantial
low-energy tail. The real parts change sign near the corresponding
peaks in the imaginary parts, consistent with causality.
}
\label{susceptibility}
\end{figure}

We first examine the lowest-order $N h$ and $\Delta h$ susceptibilities,
$\chi_N$ and $\chi_\Delta$, at normal nuclear matter density
$\rho=0.17\,\mathrm{fm}^{-3}$.
Figure~\ref{susceptibility} displays the lowest-order $N h$ and
$\Delta h$ susceptibilities at normal nuclear matter density for a
momentum transfer of $q=3m_\pi$.
The $N h$ susceptibility exhibits a response concentrated in the low-energy region, while the $\Delta h$ susceptibility provides a slightly larger response in the high-energy region owing to the strong $\pi N\Delta$ coupling. Because the full $\Delta h$ Lindhard function is evaluated by carrying out the Fermi-sea integration explicitly, the $\Delta h$ strength is
distributed over a broad energy range rather than being concentrated at a
single excitation energy.
Consequently, the $\Delta h$ excitation extends well into the low-energy
region, where it mixes strongly with the $N h$ excitation.
This mixing plays a crucial role in generating the collective pion modes
discussed below.

The qualitative features of the response function are insensitive to the choice of the $\Delta$ width.
The width primarily smooths the spectral distribution, whereas the
dominant enhancement of the low-energy response originates from the
full $\Delta h$ Lindhard function obtained by explicitly carrying out the
Fermi-sea integration.
Even in the limit of a vanishing width,
$\Gamma_0\rightarrow0$,
the imaginary part of the $\Delta h$ susceptibility remains finite over
the kinematically allowed energy interval for the $\D h$ excitation
$[\omega_{\min},\omega_{\max}]$, where
\beq
\omega_{\max,\min}
=
\omega_\Delta
+\frac{(k_F\pm q)^2}{2M_\Delta}
-\frac{k_F^2}{2M_N}~.
\eeq
For $q=3\,m_\pi$, the corresponding energy range is
$\omega_{\min}=1.8m_\pi$
to
$\omega_{\max}=3.2\,m_\pi$.

Figure~\ref{response} shows the imaginary part of the response function
at normal nuclear matter density for a momentum transfer
$q=3\,m_\pi$.
The black curve represents the lowest-order polarization function,
$-\mathrm{Im}\,\Pi$.
The blue, red, and green curves show the RPA response,
$-\mathrm{Im}\,R$,
including both $Nh$ and $\Delta h$ correlations for
different values of the Landau-Migdal parameter $g'$.

\begin{figure}[h]
\centering
\vspace{+1.5mm}
\includegraphics[width=\linewidth]{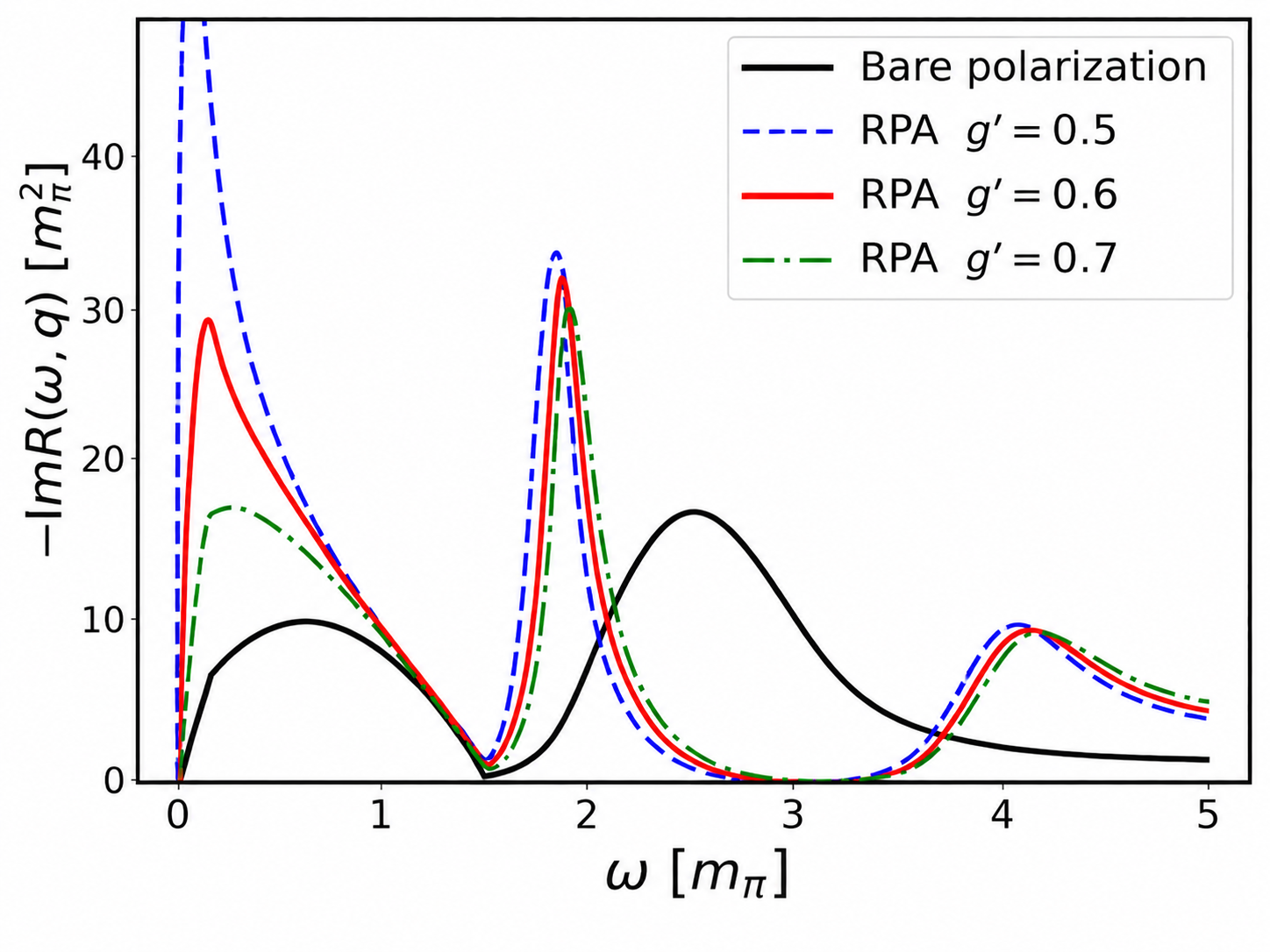}
\vspace{-4.5mm}
\caption{
Imaginary part of the RPA response function,
$-\mathrm{Im}\,R(\omega,q)$, at normal nuclear matter
density for $q=3m_\pi$.
The black curve represents the lowest-order polarization
function, $-\mathrm{Im}\,\Pi(\omega,q)$, while the blue,
red, and green curves show the RPA responses including
both $Nh$ and $\Delta h$ excitations for different
values of the Landau-Migdal parameter $g'$.
A pronounced low-energy collective pion mode emerges
around $\omega\sim0.2m_\pi$ and becomes increasingly
enhanced as $g'$ decreases.
A second collective resonance appears near
$\omega\simeq1.8m_\pi$, originating predominantly
from the $\Delta h$ channel.
}
\label{response}
\end{figure}

The large magnitude and broad energy distribution of
$\chi_\Delta$, arising from the full $\Delta h$ Lindhard function,
strongly modify the RPA denominator,
$C_{\pi\Delta}=1-D(q,\omega)\Pi(q,\omega)$.
This leads to collective modes through the coherent coupling of
$N h$ and $\Delta h$ excitations.
As the Landau-Migdal parameter $g'$ is reduced,
the attractive pion-mediated interaction becomes stronger, thereby enhancing the 
collective response.

A pronounced collective mode emerges in the low-energy region around $\omega \sim 0.2\,m_\pi$. This mode is absent in the lowest-order response and develops only after the full RPA summation of both the $N h$ and $\D h$ excitations. Its strength increases rapidly as the denominator approaches zero, indicating the formation of a strongly collective pion excitation. The behavior closely resembles the precursor phenomenon discussed in pion-condensation studies, although the present system remains far from the condensation point. 

A second collective structure appears near $\omega \simeq 1.8 \,m_\pi$, below the $\Delta h$ excitation region. This peak structure has been discussed by Chanfray et al. \cite{Chanfray1984}; however, their resonance appears at a somewhat
lower excitation energy because the coupling to the $Nh$ excitations
was not treated explicitly.  This resonance originates primarily from the strong $\pi N\Delta$ coupling and the collective nature of the $\Delta h$ excitations. In contrast to the low-energy mode, however, the detailed properties of this high-energy structure are expected to depend on the in-medium $\Delta$ self-energy.  This delta-hole collective peak becomes sharper as $\Gamma_0$ is reduced. 

A repulsive $\Delta h$ collective mode is also expected at higher excitation energies. Since its excitation energy lies well beyond the region of present interest, we do not discuss this mode further.

We therefore regard the low-energy collective pion mode as the most robust prediction of the present framework. At the same time, the appearance of the high-energy resonance demonstrates that $\Delta h$ correlations play a much more significant role in the nuclear spin-isospin response than previously recognized.  Although a detailed comparison has not yet been made, similar features of the collective response functions appear to have been observed in the high-energy $(^3\mathrm{He},t)$ reaction \cite{Oset1989}.

\begin{figure}[t]
\centering
\includegraphics[width=\linewidth]{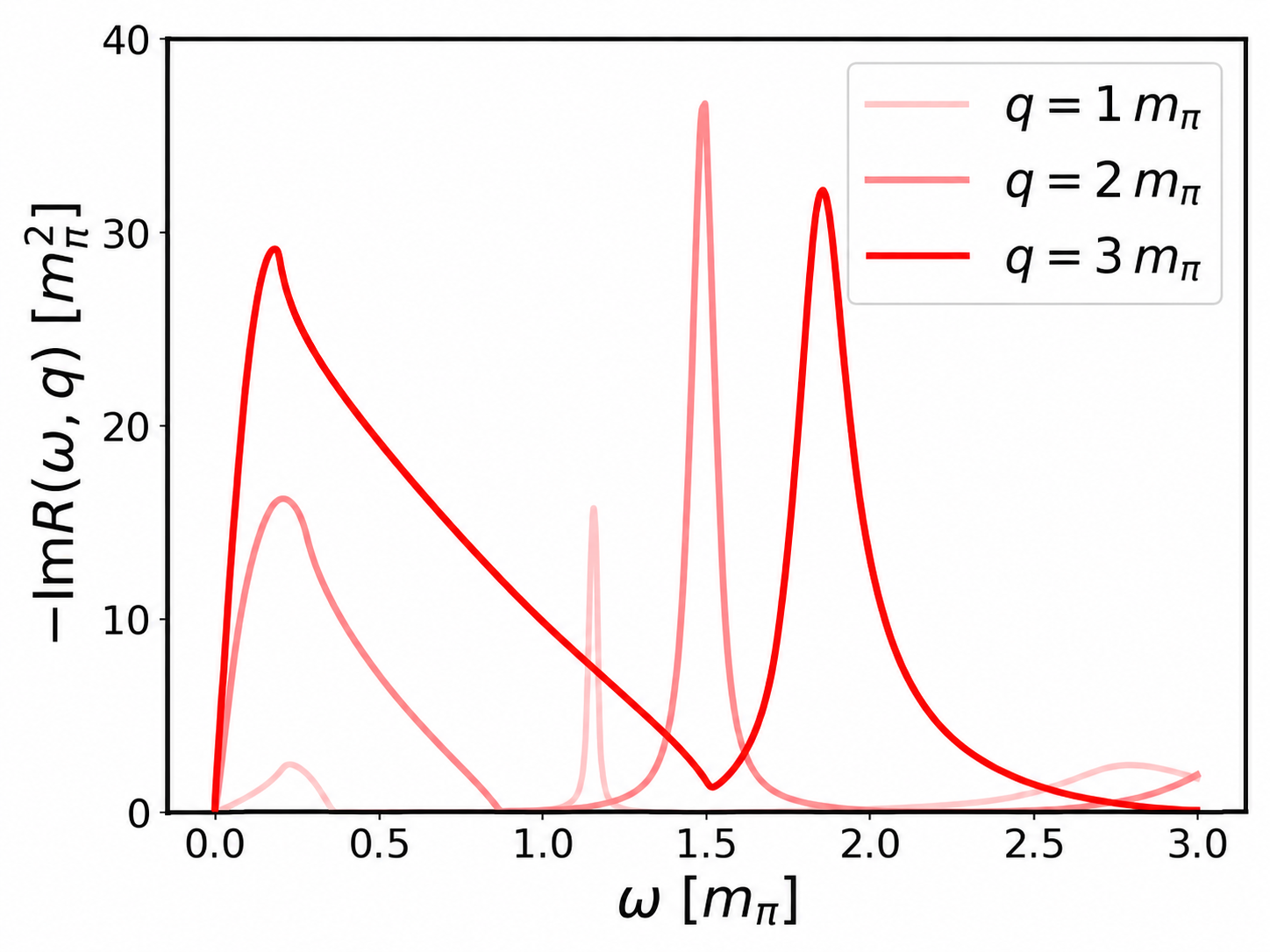}
\vspace{-3mm}
\caption{
Imaginary part of the RPA response function,
$-\mathrm{Im}\,R(\omega,q)$, at normal nuclear matter
density for momentum transfers $q=1,2,3m_\pi$.
The Landau-Migdal parameter is fixed at $g'=0.6$.
For $q=1m_\pi$ (red curve), the low-energy collective
pion mode around $\omega\sim0.2m_\pi$ is strongly
suppressed, whereas a sharp collective $\Delta h$
resonance appears near $\omega\sim1.1m_\pi$.
For $q=2m_\pi$ (blue curve) and $q=3m_\pi$
(brown curve), the low-energy collective pion mode
becomes increasingly enhanced with momentum transfer.
The high-energy collective mode appears below the
$\Delta$--$N$ mass difference and shifts toward
higher excitation energies as $q$ increases.
}
\label{responseq}
\end{figure}

In Fig.~\ref{responseq}, we compare the response functions, $-{\rm Im}\, R$, for different momentum transfers at fixed $g'=0.6$. The low-energy collective pion mode is strongly suppressed at $q=1m_\pi$ and grows in strength as the momentum transfer increases. This is because the interaction among $Nh$ states is expressed as Eq.\,(\ref{interaction}) and at small $q$ the attractive and repulsive contributions cancel each other, while at large $q$, the interaction becomes largely attractive. At $q=2m_\pi$, both the $Nh$ and $\Delta h$ components are substantially enhanced, while at $q=3m_\pi$ the low-energy collective structure becomes particularly pronounced.  The response functions depend largely on the momentum transfer, and it is very important to study the response functions as functions of the momentum transfer $q$.  We mention here that the similar spectra are presented by Dmitriev and Suzuki, but their choice of $g'$ is large and the enhancement observed here is not predicted \cite{DmitrievSuzuki1985}.

We now discuss the difference in pionic correlations between
proton-neutron ($pn$) pairs and like-nucleon pairs ($pp$ and $nn$).
The dominant two-particle-two-hole excitation in the pion channel
arises from the coupled $^3S_1$--$^3D_1$ channel, where the nucleon pair
must have isospin $T=0$ because of the Pauli principle.
Consequently, pionic correlations are strong for $pn$ pairs but are
strongly suppressed for $pp$ and $nn$ pairs.
This picture is consistent with modern studies of tensor and
short-range correlations, which have demonstrated the dominance of
correlated $pn$ pairs over $pp$ pairs in nuclei \cite{Alvioli2008}.
Within the present RPA framework, this suppression may be described
phenomenologically by a larger effective value of the Landau-Migdal
parameter $g'$ for like-nucleon pairs.
We therefore expect the hybrid pion-delta collective modes to appear
predominantly in the charged-pion channels, whereas they should be
strongly suppressed in the neutral-pion channel.

The collective pion-delta modes predicted here are expected to be
observed in high-momentum-transfer reactions probing the
longitudinal spin-isospin response of nuclei.
Proton-induced charged-pion production reactions,
such as $(p,p'\pi^+)$,
provide a particularly promising probe because they selectively
populate excitations carrying large momentum transfer and pion
quantum numbers.
Complementary information may also be obtained from reactions
such as ($^3$He,$t$) and $(p,n)$, which are sensitive to the
spin-isospin response.

A particularly important feature of the present framework is that the
same pion-delta collective function governs both the response
function and the RPA correlation energy.
Consequently, the collective pion modes observable in
high-momentum-transfer $(p,p'\pi^\pm)$ reactions are directly linked to
the pion-delta correlations responsible for an attractive
contribution to the nuclear binding energy.
The present work therefore establishes an explicit connection between
an experimentally accessible collective excitation and the microscopic
origin of the RPA correlation energy.
The implications for nuclear binding will be presented elsewhere.

The same pion-delta dynamics responsible for the collective response also contributes to the ground-state energy through the Gell-Mann--Brueckner expression~\cite{Gell-Mann1957, Fetter2003},
\beq
\label{RPAenergy}
E_{\rm corr}^{\rm RPA} = \frac{1}{2} \int \frac{d^4q}{(2\pi)^4} \left[ \ln(1 - D\Pi) + D\Pi \right].
\eeq
The connection between collective excitations and nuclear binding is particularly transparent because both quantities are governed by the same denominator, the collective pion--delta function $\mathcal{C}_{\pi\Delta}$.
An enhancement of the response function therefore implies a suppression of this denominator, which simultaneously increases the attractive contribution to the RPA correlation energy. The collective pion modes discussed above are thus directly linked to the microscopic mechanism generating nuclear binding.

\begin{table}[h!]
\centering
\caption{RPA correlation energy per nucleon,
$E_{\mathrm{corr}}^{\mathrm{RPA}}/A$ (MeV),
at normal nuclear matter density as a function
of the Landau-Migdal parameter $g'$.
The $Nh$-only calculation includes the $Nh$
susceptibility $\chi_N$, while the $Nh+\Delta h$
calculation includes both $\chi_N$ and $\chi_\Delta$,
with the latter evaluated using the full $\Delta h$
Lindhard function of Eq.~(19).
The $\Delta$ width is set to zero in both the
direct and crossed terms.}
\label{tab:rpaenergy}
\begin{ruledtabular}
\begin{tabular}{ccc}
$g'$ & $Nh$ only & $Nh+\Delta h$ \\
\hline
0.45 & $-13.1$ & $-56.6$ \\
0.50 & $-9.9$  & $-40.7$ \\
0.55 & $-7.3$  & $-29.5$ \\
0.60 & $-5.3$  & $-21.1$ \\
0.65 & $-3.6$  & $-14.6$ \\
0.70 & $-2.4$  & $-9.8$  \\
\hline
0.90 & $-0.5$  & $-4.2$  \\
\end{tabular}
\end{ruledtabular}
\end{table}

We evaluate Eq.~(\ref{RPAenergy}) after Wick rotation to the Euclidean frequency axis. Table~\ref{tab:rpaenergy} compares the RPA correlation energy per nucleon calculated without $\Delta h$ excitations to that including the corrected $\Delta h$ susceptibility of Eq.~(\ref{chid}). In the latter calculation, we set the $\Delta$ width to zero, as taking a finite width yields a negligible contribution.

Including the $\Delta h$ channel enhances the attractive correlation energy across all displayed values of $g'$. At $g'=0.45$, the correlation energy changes from $-13.1$~MeV in the $Nh$-only calculation to $-56.6$~MeV with $\Delta h$ excitations; at $g'=0.70$, the corresponding values are $-2.4$ and $-9.8$~MeV. Remarkably, the inclusion of the $\Delta h$ contribution increases the magnitude of the correlation energy by nearly a factor of four. The correlation energy becomes less negative as $g'$ increases, with the additional attraction provided by the $\Delta h$ channel being most pronounced at smaller values of $g'$.

Note that the correlation energies in Table~\ref{tab:rpaenergy} include only the charged-pion contributions originating from proton-neutron excitations. Excitations of like-nucleon pairs ($pp$ and $nn$) are strongly suppressed by the Pauli principle, as discussed in the context of the response function. Within the present RPA formalism, this suppression of like-pair contributions can be effectively represented by adopting a larger value of $g'$. For comparison, Table~\ref{tab:rpaenergy} also presents the case of $g'=0.90$, where the correlation energies both with and without $\Delta h$ excitations remain small.

\section{Conclusions}

We have investigated the longitudinal spin-isospin response of symmetric nuclear matter including both $N h$ and $\Delta h$ excitations within the random-phase approximation.

The central result of the present work is the emergence of collective pion-delta modes generated by the coherent coupling between $Nh$ and $\D h$ excitations. Unlike many earlier treatments in which the $\Delta h$ excitation was represented by a single excitation energy, we evaluate the full $\Delta h$ Lindhard function, including the Fermi-sea integration. This treatment generates a substantial low-energy tail in the $\Delta h$ response, leading to strong mixing between the $Nh$ and $\D h$ channels and giving rise to collective pion-delta modes.

As a result, a pronounced low-energy collective pion mode emerges at momenta of the order of a few pion masses. The strength of this mode increases as the Landau-Migdal parameter $g'$ is reduced, exhibiting the characteristic behavior expected for a precursor of pion condensation. In addition, a second collective pion-delta mode appears near the $\Delta h$ excitation region, demonstrating the important role of delta-hole correlations in the nuclear spin-isospin response.

The collective pion-delta modes predicted in the present work are expected to be accessible through high-momentum-transfer reactions probing the longitudinal spin-isospin response. High-momentum-transfer charged-pion production reactions are expected to provide the most direct probe of these collective modes.

An important feature of the present framework is that the same
collective pion-delta function,
${\cal C}_{\pi\Delta}(q,\omega)$,
governs both the response function and the RPA correlation energy.
Collective modes emerge as ${\cal C}_{\pi\Delta}$ approaches zero,
while the same suppression enhances the attractive RPA correlation
energy.
The present work therefore establishes an explicit connection
between experimentally observable collective pion excitations and
the microscopic origin of the RPA correlation energy.

\begin{acknowledgments}
The authors gratefully acknowledge financial support
from the Yamada Science Foundation.
\end{acknowledgments}

\bibliographystyle{unsrt}
\bibliography{pionbib}

\end{document}